# A generalization of Amdahl's law
# and relative conditions of parallelism

**Author**: *Gianluca Argentini*, New Technologies and Models, Riello Group, Legnago (VR), Italy. E-mail: gianluca.argentini@riellogroup.com

**Abstract**: In this work I present a generalization of Amdahl's law on the limits of a parallel implementation with many processors. In particular I establish some mathematical relations involving the number of processors and the dimension of the treated problem, and with these conditions I define, on the ground of the reachable speedup, some classes of parallelism for the implementations.
I also derive a condition for obtaining superlinear speedup.
The used mathematical technics are those of differential calculus.
I describe some examples from classical problems offered by the specialized literature on the subject.

## 1. Introduction

In the world of parallel computing or in general of high performances one of the metric more useful for evaluating the gain reachable in an implementation on many processors of a program in comparison with its serial monoprocessor version is the *speedup S* (v. PACHECO, 1997), defined as the ratio between the time $T_{ser}$ occurred for the execution of the serial program and the time $T_{par}$ occurred for its parallel version:

$$S = \frac{T_{ser}}{T_{par}} \qquad (1)$$

In this work I consider these two times as computed by a *scalability analysis* (v. GROPP, 2002) of a particular logic implementation of the problem which they refer to, and not by their measurement on a particular hardware system.
From the formula (1) one obtain the Amdahl's Law (v. AMDAHL, 1967) by mean of the concept of parallelizable fraction $f$ of a particular parallel implementation, that is the percentage of statements that are executable at the same time on many processors:

$$S = \frac{p}{f(1-p)+p} \qquad (2)$$

where $p$ is the number of used processors and $0 < f \leq 1$. The optimal case, that is when $f = 1$, provides for $S$ a value equal to $p$. The $S$ is an increasing function respect to the variable $p$, and for $p$ tending to infinity we obtain the limit

$$\frac{1}{1-f} \qquad (3)$$

that expresses the Amdahl's law, from which one observe that the speedup admits a superior limitation determined by the used code, even if the number of processors *p* is very high.

In figure 1 is drawn the graphic of speedup for a code with $f = 0.8$; the asymptotic limit is 5:

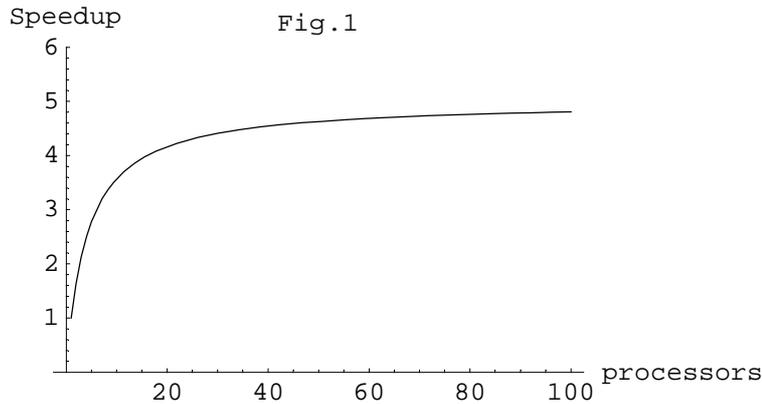

But the limitation imposed from this law can be overtaken, even in a a large measure, if one thinks that in formula (2) is not explicitly present the parameter that expresses the *dimension* of the considered problem, that is the number *n* of data given as input. A first reexamination of the Amdahl's law (v. BENNER - GUSTAFSON - MONTRY, 1988) shows that, in several real situations, when the number of processors increases, a corresponding opportune increase of the number of treated data provides a speedup much bigger than that imposed by (2). Even in (PACHECO, 1997) there is a brief but illuminating discussion of this possibility.

The purpose of this work is to explain from a mathematical point of view how it is possibile to obtain speedup values much higher than those estimated from Amdahl's law, and to classify the parallel implementations on the ground of the speedup obtained by various combinations of the parameters *n* and *p*.

**2. Generalization of Amdahl's law**

I will refer to an implementation of a given problem, that is to a triad (Problem, Program, System) constituted by a problem, for example the multiplication of a matrix for a vector, by a program that accepts the problem's data as input and by an operative environment, to be intended as hardware as software, into which the program runs. The problem gives the dimension *n* of the implementation; the program, adapting itself even to the hardware that is used, gives the number *p* of processors. In accordance with the most part of the literature on the high performances, for simplicity I identify the number of processors with the number of independent used processes, equal copies of the considered program. For example, in the codes that use the parallelization library MPI (see MPI FORUM) the start of a parallel program permits to specify the number of processes, that is of independent copies of the program itself, that communicate each other during the execution. Also I use the condition that the processors are all of the same type and that the operative environment is homogeneous respect to the used software (p.e. compilers, libraries).

The first consideration is that the parameter *f* of Amdahl's Law can depend from the used implementation, that is from the code instructions of the program used for treating the problem, and hence in general it will be a function of *p* and *n*:

$$f = f(p, n) \tag{4}$$

If we introduce (4) into (2) and derive partially respect to *p*, we obtain

$$\frac{\partial S}{\partial p} = \frac{\frac{\partial f}{\partial p} p^2 - \frac{\partial f}{\partial p} p + f}{(f + (1-f)p)^2} \tag{5}$$

One realist and desirable condition in the parallel implementations for high performances is that the speedup grows when the numbers of used processors increases, hence we impose the following condition:

$$\frac{\partial S}{\partial p} \geq 0 \tag{6}$$

and from the fact that the denominator in (5) is always positive it follows that

$$\frac{\partial f}{\partial p} p^2 - \frac{\partial f}{\partial p} p + f \geq 0 \tag{7}$$

Since $f > 0$, (7) can be written in this way:

$$1 + (p^2 - p) \frac{\frac{\partial f}{\partial p}}{f} \geq 0 \tag{8}$$

Also, since $p > 1$ in a parallel implementation, follows that $p^2 - p > 0$,

and remembering that $\frac{\frac{\partial f}{\partial p}}{f} = \frac{\partial (\text{Log } f)}{\partial p}$, from (8) one obtains:

$$\frac{1}{(p^2 - p)} + \frac{\partial (\text{Log } f)}{\partial p} \geq 0 \tag{9}$$

Indefinitely integrating the first addendum respect to the variable *p*, (9) is equivalent to

$$\frac{\partial}{\partial p} \left( \text{Log } \frac{p-1}{p} + g(n) + \text{Log } f \right) \geq 0 \tag{10}$$

where *g*(*n*) is an arbitrary function of *n*, that is the constant of integration respect to *p*. Indicating with $F(p, n)$ the sum of the two logarithmic expressions, one can write

$$\text{Log } f = F(p, n) - \text{Log } \frac{p-1}{p} = F(p, n) + \text{Log } \frac{p}{p-1} \tag{11}$$

from which follows that:

$$f = \frac{p}{p-1} \text{Exp}(F(p, n)) \qquad (12)$$

We call the function *F exponent of parallelism*.

We see now two *F* 's properties useful for the subsequent discussion. First of all from (10) follow

$$\frac{\partial F}{\partial p} \geq 0 \qquad (13)$$

hence *F* is an increasing function respect to *p*. Also, being by definition $f \leq 1$, from (12) one obtains the following condition:

$$F(p, n) \leq \text{Log}\frac{p-1}{p} < 0 \qquad (14)$$

Now we'll do some reasonings about the relation between *F* and the parallelism of the relative implementation.

In the first place we can notice that it's realistic assuming that the dimension *n* of a treated problem is greater of the number *p* of processors used in the corresponding parallel implementation; on the contrary we have a waste of hardware resources. Also, as explicited p.e. in (PACHECO, 1997) and (BENNER - GUSTAFSON - MONTRY, 1988), it's realistic assuming that as the number of used processors increases, the problem dimension can increase too, otherwise the concept of parallelism's performance and the research of a good implementation itself could have no meaning. Under these ipothesis from (12) there are two possibilities:

*A)* if *p* becomes very large, the ratio $\frac{n}{p}$ tends to a finite limit greater than 0; this is the case for example when $n = k\, p$, where *k* is a constant, even large; in this situation the two possible meaningful alternatives are, remembering the conditions (13) and (14):

*a')* $F(p, n)$ tends to 0, and from this follows that *f* tends to 1; hence the parallel implementation allows an unlimited speedup when *p* increases, under the condition that the growth of problem's dimension *n* is asymptotic to the processors number;

*a'')* $F(p, n)$ tends to a finite limit smaller than 0, and from this follows that *f* tends to a value $f_0$ where $0 < f_0 < 1$; in this case, when the processors number increases, the parallel implementation gives a behaviour that conforms to Amdahl's law, and the speedup is equal to $\frac{1}{1-f_0}$;

*B)* if *p* becomes very large, the ratio $\frac{n}{p}$ tends to $+\infty$; this is the case for example when $n = p \text{Log}\, p$; in this situation the two possible meaningful alternatives are:

*b')* $F(p, n)$ tends to 0, and from this follows that *f* tends to 1; hence the parallel implementation allows an unlimited speedup when *p* increases, and the problem dimension can now increases with a very high velocity respect to the processors number; hence this situation is optimal and a real case is given for example in (PACHECO, 1997) by a parallel implementation of the numerical integration with the trapezoidal rule;

*b'')* $F(p, n)$ tends to a finite limit smaller than 0, and from this follows that *f* tends to a value $f_0$

where $0 < f_0 < 1$; hence in this case too, when the processors number increases, the parallel implementation gives a behaviour that conforms to Amdahl's law, but the problem dimension can now increase in a way not asymptotic to the processors number.

I consider not very realistic or at least not meaningful for an analysis of their parallelism the situations where the exponent of parallelism tends to -∞, in which case f tends to 0, the speedup becomes 1 and there is resources's waste, or whose for which the ratio $\frac{n}{p}$ tends to 0.

From the previous considerations we can get the following *theorem-definition*, which generalizes the Amdahl's law and establishes the parallelism type for a given implementation:

> In a parallel implementation, *n* is the dimension of the relative problem, *p* the processors number, *F* (*p,n*) the function exponent of parallelism.
>
> If $n = g(p)$ is an increasing function of *p* such that the ratio $\frac{g(p)}{p}$ tends to +∞ and *F* (*p*, *g(p)*) tends to 0 for *p* tending to +∞, than the implementation is **strongly parallel**.
>
> If *F* (*p*, *g(p)*) tends to 0 for *p* tending to +∞ only for *g(p)* increasing function of *p* such that the ratio $\frac{g(p)}{p}$ tends to a finite limit, than the implementation is **weakly parallel**.
>
> If for every *g* (*p*) increasing function of *p* the function *F* (*p*, *g(p)*) tends to a limit smaller than zero for *p* tending to +∞, than the implementation is **Amdahl-like parallel**.

### 3. Considerations and examples

When we apply the preceding classification we meet two problems: the first is how to calculate the function exponent of parallelism, and the second is how to demonstrate if exist the right *g* (*p*) function. The former depends upon the used implementation and it can be solved by mean of a scalability analysis (see GROPP, 2002), and in the following considerations I'll try to explain how it is possible to obtain some useful information; the latter is a problem of mathematical kind which can be tackled by mean of the methods of differential calculus.

First of all we can notice that in (12) we are interested to know when, for large enough values of *p*, the exponential values are near to 1, and hence if we write the polynomial series of *Exp(F)* respect to the argument *F* we can assert that the following is a good approximation:

$$f = \frac{p}{p-1}\left(1 + F + \frac{F^2}{2}\right) \quad (15)$$

Using this expression of *f* into (2) one can obtain the following formula:

$$F = -1 \pm \sqrt{1 - \frac{2}{S}} \quad (16)$$

where the interesting case is that with the + sign, because it is useful for examine the closeness of $F$ to 0.

In (1) we can consider $Tpar = Tpar(p, n)$. Also we can establish, as a good *approximation* in general and optimal in the case when all the system's processors are of the same kind, that $Tser = Tpar(1, n)$, i.e. the time spent in a serial implementation of a program can be approximated with the time spent in the corrisponding parallel implementation executed in a unique processor. From (1) and (16) one obtain the following parallelism condition:

$$\mathrm{Tpar}(p, n) \leq \frac{\mathrm{Tpar}(1, n)}{2} \qquad (17)$$

which can be interpretated as a *minimal condition of parallelism* for the warranty of an advantageous speedup.

We now examine some example in which using the relation

$$F = -1 + \sqrt{1 - \frac{2 T par(p, n)}{T par(1, n)}} \qquad (18)$$

we can obtain the informations given in the preceding generalization of Amdahl's law.

In (PACHECO, 1997) is presented a parallel implementation of the numeric computation of a definite integral by mean of the trapezoidal rule, and using a scalability analysis the following estimate is obtained:

$$T par(p, n) = a \frac{n}{p} + b \operatorname{Log} p$$

whee $a$ and $b$ are two positive constants depending from the used operative environment. Hence we have

$$\frac{\mathrm{T} par(p, n)}{\mathrm{T} par(1, n)} = \frac{a n + b p \operatorname{Log} p}{a n p}$$

from which we see, using the rules of calculus, that if one consider for example $n = p^2$, increasing function respect to $p$, the limit of the ratio when $p$ indefinitely grows is 0, and from (18) one obtain that $F$ tends to 0. Hence on the ground of the preceding classification, the implementation is *strongly parallel*, and this fact is in accordance with the scalability argumentations developed from Pacheco, from which if the increment of $n$ is opportunity guided by that of $p$, the speedup remains high. One can see that from the preceding ratio for applying the classification rule it is sufficient to use the function $n = p \operatorname{Log} p$, as reported by Pacheco. If one uses the function $n = p$, the limit of $F$ when $p$ increases is still 0, but since the ratio $\frac{n}{p}$ tends to a finite limit the implementation, already classified as strongly parallel, reveals in such conditions a weakly parallel behaviour. This fact shows, in accordance with (BENNER -GUSTAFSON-MONTRY, 1988), the importance of a convenient growth function $n = g(p)$ in a parallel implementation which would have as aim a high speedup.

Also it can be notice that keeping fixed the dimension $n$ of the problem, when the processors number increases the preceding ratio increases in an unbounded manner, hence the parallelism is no more advantageous, the (18) is not applicable, *S* tends to zero and hence the Amdahl's law (3) turns out not correct. This fact suggests the hypotesis that in the set of possible implementations *the strongly parallel implementations are those that turn out advantageous when the problem's dimension grows in a suitably considerable way respect to the used processors number.*
In contrast with the Fig.1, we present in the following Fig.2 the graphics of the speedup for the trapezoidal rule respectively in the case $n = p^2$, $n = p\, Log\, p$ e $n = p$:

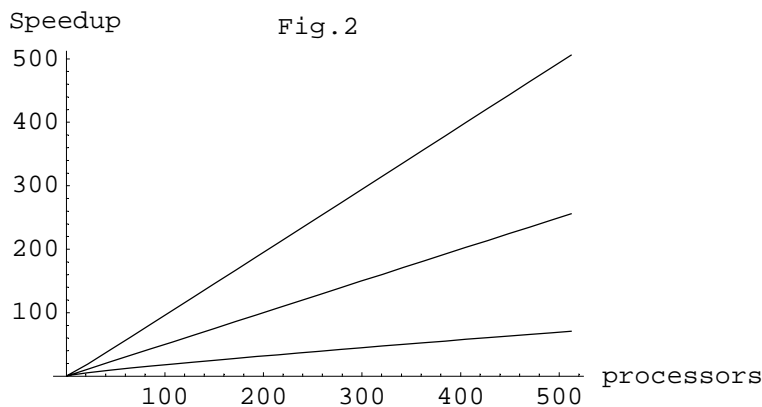

from which one can see that the speedup improves when the derivative respect to the processors number of the problem dimension grows.

In (CORMEN-LEISERSON-RIVEST, 1990) is presented a parallel implementation for the calculation of a Fast Fourier Transform which has the aim of keeping an asymptotic execution time $Tpar = A\, Log\, (n)$, where *A* is a constant, *n* is the input dimension and the logarithm is in base 2. The corresponding serial implementation shows an asymptotic time $Tser = B\, n\, Log\, (n)$. The constants *A* and *B* depend from the used operative environment. The parallel procedure is obtained by mean of a necessary configuration of *n* combinatorial elements, which are responsable of the operations of addition, multiplication and intercommunication of partial results. If one use *p* processors, each of these assembles *k* combinatorial elements, it is $n = k\, p$, and hence

$$\frac{Tpar}{Tser} = \frac{A}{B\, n} = \frac{A}{B\, k\, p}$$

The conclusion is that the function *F* tends to 0 for *p* tending to +∞, and this under the condition that the problem dimension grows linearly respect to the processors number. Hence on the ground of the preceding classification, the parallel implementation proposed for the FFT is *weakly parallel.*
In this example if one keeps constant the problem dimension *n*, increasing even indefinitely the processors (p.e. identifying each of them with a single combinatorial element, hence imposing *k* = 1), the ratio of the execution times is constant too, and therefore in this case the Amdahl's law can be applied. This consideration suggests that in the set of possible implementations *the weakly parallel implementations are those for which, keeping constant the relative problem's dimension, the Amdahl's law is applicable.*

In (GROPP, 2002) is presented an implementation for multiplying a $n \times n$ matrix with a vector of $n$ components. The scalability analysis gives the following estimate:

$$Tpar(p, n) = \frac{a(2n^2 - n)}{p} + b(n^2 + n)$$

where $a$ and $b$ are two positive constants depending from the used operative environment. Hence one obtain

$$\frac{Tpar(p, n)}{Tpar(1, n)} = \frac{a(2n^2 - n) + pb(n^2 + n)}{p[a(2n^2 - n) + b(n^2 + n)]} = \frac{bpn^2 + 2an^2 + bpn - an}{(2a + b)pn^2 + (-a + b)pn}$$

from which we see, using the rules of calculus, that if one considers a whatever increasing function $n = g(p)$ such that $\frac{g(p)}{p}$ tends to $+\infty$ or not, the limit of the ratio when $p$ grows is

$$\frac{b}{(2a + b)}$$

Supposing that this quantity satisfies the minimal condition of parallelism (17), the limit of the function $F$ given by (18) is finite and negative, therefore the rule of classification previously enunciated specifies that the considered implementation is *Amdahl-like parallel*. In (GROPP, 2002) is shown that the parameter $b$ is the number of microseconds spent by a specific hardware for communicate beetwen two processes a floating-point quantity, while $a$ is the time spent for the execution of a floating-point operation. The (17) should imply the condition $b \leq 2a$, not very realistic with the present hardware, but the improvements on the communications technology beetwen processes could render it reliable in the future. Imposing $b = 2a$, from (18) we have $F = -1$, hence from (15) we obtain for $f$ the limit 0.5, therefore the asymptotic value of the speedup is 2, that is near to the value 2.3 obtainable directly from (1); in the following Fig.3 I present the graphics of the speedup in the case $n = p^2$, $n = p \, Log \, p$ and $n = p$, from which it is clear that the increasing of the dimension problem respect to the used processors number practically hasn't influence, on the contrary of what happens in a strongly parallel implementation:

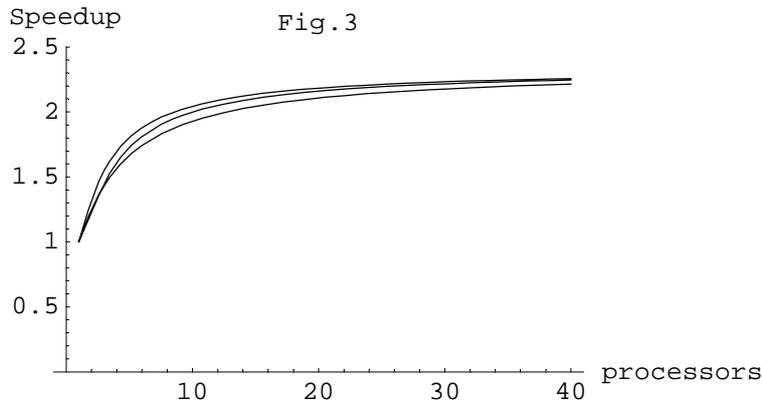

### 4. Superlinear speedup
In some real situations a superlinear speedup is been registered, that is for some values of $p$ and *n* the experimental value of *S* is resulted greater then *p*. A necessary condition for this situation is that in (2) the function *f* assumes values greater than 1, and precisely

$$f > \frac{p}{p-1} \tag{19}$$

This doesn't agree fullfully with the initial definition of *f* as parallelizable fraction of a code, and hence smaller or equal to 1. Now we see how the superlinear speedup can be explained, in the mathematical proposed model, in a fashion coherent with the original definition of *f*.
From (2) and (12) the superlinear speedup imposes that

$$\frac{p}{p-1} \mathrm{Exp}(F(p,n))(1-p) + p < 1 \tag{20}$$

and hence it must be

$$\mathrm{Exp}(F(p,n)) > 1 - \frac{1}{p} \tag{21}$$

therefore one obtains the condition

$$F(p,n) > \mathrm{Log}\left(1 - \frac{1}{p}\right) \tag{22}$$

First of all we can notice that the argument of logarithm is smaller than 1, therefore the (22) can be satisfied for values of *F* smaller than 0 too, and hence the proposed description is coherent with the originary condition $f \leq 1$.
We'll use the following approximations, that are good for small values of the argument *x*:

$$\mathrm{Log}(1+x) = x - \frac{x^2}{2} \tag{23}$$

$$\sqrt{1+x} = 1 + \frac{x}{2} - \frac{x^2}{8} \tag{24}$$

In the case of a superlinear implementation the quantity under logarithm in (22) in general is small (mathematically it would sufficient $p = 10$), therefore applying (23), from (22) and (16) one obtains

$$F(p,n) > -\frac{1}{2p^2} - \frac{1}{p} \tag{25}$$

that can be considered as a *superlinear condition* of an implementation characterized by an exponent of parallelism *F (p,n)*.
The figure 4 presents the graphic of the second member of (25), that represents the inferior limit which must be respected by the function *F (p,n)*:

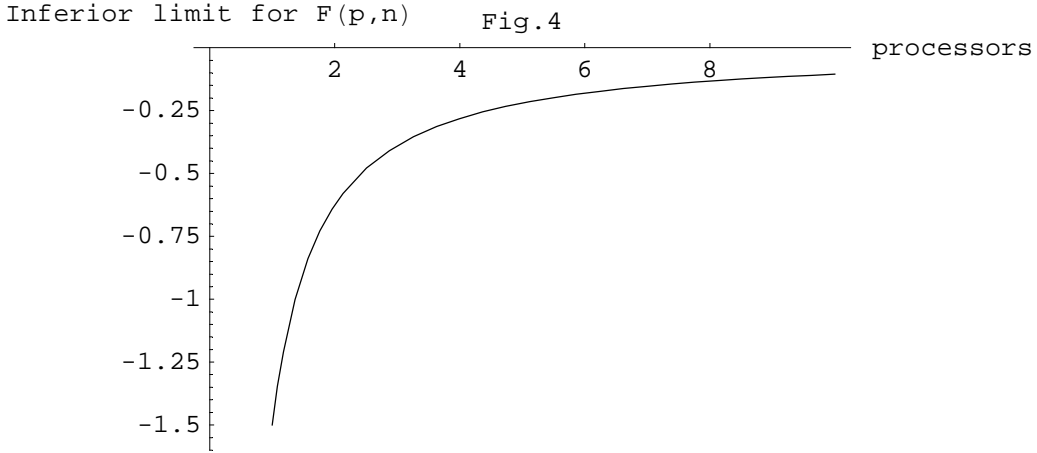

In such conditions the argument under square root in (16) is near to 1, hence from (24) one obtains

$$\frac{1}{2S^2} + \frac{1}{S} < \frac{1}{2p^2} + \frac{1}{p} \qquad (26)$$

condition which is trivialy satisfied by the classical condition of superlinearity $S > p$. Therefore (25) and (26) extend the notion of superlinearity, and this in a coherent fashion with the original condition $f \leq 1$.

As an example of application of (25) or (26), we consider the parallel implementation of the Fast Fourier Transform already mentioned. In this case one has, as upon reported,

$$\frac{1}{S} = \frac{A}{Bn}$$

We suppose for simplicity that $n$, as happens in real applications, is sufficiently large so that in (26) one can disregard the quadratic term at first member. Denoting by $C$ the quantity $\frac{A}{B}$, that depends from the hardware and software environment, using (26) and solving the disequality respect to the variable $p$ one obtains

$$p < \frac{1}{2C}\left(n + \sqrt{n^2 + 2Cn}\right) \qquad (27)$$

hence in this example the superlinear speedup is possible only if the processors number is upper limited by a relation that involves the problem dimension. In (CAVAZZONI-CHIAROTTI, 2001) is reported the experimental observation of superlinear speedup in an implementation that uses in a great and sofisticated manner many parallelized FFT on a system Cray T3E with Fortran 90 as compiler: the processors region which presents superlinearity is upper limited, and the phenomenon is due to the effects of hardware and software caches, which in the preceding mathematical schema are pointed out by the presence of the constant $C$.

## 5. Conclusions

In this work I proposed a mathematical interpretation of some experimental results and of some theoretical digressions reported in the literature on the possible limits and performances of parallel computing. I proposed a generalization of Amdahl's law on the possible speedup obtainable in a parallel implementation. In particular I have presented some sufficient conditions in order that, in a given implementation, the speedup could indefinitely grow when the dimension of the analyzed problem increases as consequence of the growing of the used processors number. By mean of these conditions I have defined three classes into which the parallel implementations can be classified, and the discriminant agent is offered by the relation of growth of problem's dimension respect to the used processors number. Also I have proposed a condition of superlinear speedup that is coherent with the original definition of parallelizable fraction of an implementation. Some concrete examples are been presented to illustrate the formulated mathematical description, which in particular show that the obtained conditions have some constants depending from the hardware and software environment where the parallel implementation is executed.

Possible further developments can regard the formulation of an algorithm for calculate the function exponent of parallelism of an implementation, and an extension of the mathematical model to hardware architectures with non homogeneous processors.